# Efficient Fizeau Drag from Dirac electrons in monolayer graphene

Wenyu Zhao[1]†, Sihan Zhao[1]†, Hongyuan Li[1,2,3], Sheng Wang[1,3], Shaoxin Wang[1], M. Iqbal Bakti Utama[1,3,4], Salman Kahn[1,3], Yue Jiang[1,5], Xiao Xiao[1,5], SeokJae Yoo[1], Kenji Watanabe[6], Takashi Taniguchi[7], Alex Zettl[1,3,8], Feng Wang[1,3,8]*

[1] Department of Physics, University of California at Berkeley, Berkeley, California 94720, United States.

[2] Graduate Group in Applied Science and Technology, University of California at Berkeley, Berkeley, California 94720, United States.

[3] Materials Science Division, Lawrence Berkeley National Laboratory, Berkeley, California, United States.

[4] Department of Materials Science and Engineering, University of California at Berkeley, Berkeley, California 94720, United States.

[5] Department of Physics, Chinese University of Hong Kong, Hong Kong 999077, China.

[6] Research Center for Functional Materials, National Institute for Materials Science, 1-1 Namiki, Tsukuba, 305-0044, Japan.

[7] International Center for Materials Nanoarchitectonics, National Institute for Materials Science, 1-1 Namiki, Tsukuba, 305-0044, Japan.

[8] Kavli Energy NanoSciences Institute at University of California Berkeley and Lawrence Berkeley National Laboratory, Berkeley, California 94720, United States.

† These authors contributed equally to this work

\* Correspondence to: fengwang76@berkeley.edu

Abstract：

Fizeau demonstrated in 1850 that the speed of light can be modified when it is propagating in moving media[1]. Can we achieve such control of the light speed efficiently with a fast-moving electron media by passing electrical current? Because the strong electromagnetic coupling between the electron and light leads to the collective excitation of plasmon polaritons, it will manifest as the plasmonic Doppler effect. Experimental observation of the plasmonic Doppler effect in electronic system has been challenge because the plasmon propagation speed is much faster than the electron drift velocity in conventional noble metals. Here, we report direct observation of Fizeau drag of plasmon polaritons in strongly biased graphene by exploiting the high electron mobility and the slow plasmon propagation of massless Dirac electrons. The large bias current in graphene creates a fast drifting Dirac electron medium hosting the plasmon polariton. It results in nonreciprocal plasmon propagation, where plasmons moving with the drifting electron media propagate at an enhanced speed. We measure the Doppler-shifted plasmon wavelength using a cryogenic near-field infrared nanoscopy, which directly images the plasmon polariton mode in the biased graphene at low temperature. We observe a plasmon wavelength difference up to 3.6% between plasmon moving along and against the drifting electron media. Our findings on the plasmonic Doppler effect open new opportunities for electrical control of non-reciprocal surface plasmon polaritons in nonequilibrium systems.

**Main text:**

Surface plasmon polariton (SPP), coupled electromagnetic and electron oscillation mode, has the unique capability to confine and manipulate light at subwavelength scale[2-5]. The strong light-matter interaction enhanced by the plasmon plays a key role in nanophotonics[6], ranging from nanoscale nonlinear optics[7], quantum optics[8], to flat optics[9]. Electrical control of the plasmon polariton is highly desirable in such applications. An intriguing possibility for plasmon control is through the Fizeau drag[1], where the moving electron media modulate the propagation speed of the plasmon polariton. This can be viewed as a plasmonic Doppler effect, where counter-propagating plasmons can have different speed that depends on the moving electron media.

The plasmonic Doppler effect is negligibly small in conventional noble metals because the plasmon velocity is more than a million times larger than the highest drift velocity achievable in those metals[3]. Recent theories predict that the two-dimensional (2D) Dirac electrons in graphene provide an ideal platform to realize strong plasmonic Doppler effect due to a combination of low carrier density, high electron mobility, and strong plasmon polariton confinement[10-17]. Compared with conventional metals, the electron drift velocity ($v_d$) in graphene is orders of magnitude higher and can reach a value of $3 \times 10^5$ m/s[18]. At the same time, the graphene plasmon features an ultra-high field confinement[14,19-22] ($\lambda_p/\lambda_0 \sim 1/150$), resulting in a plasmon group velocity ($v_p$) at around $2 \times 10^6$ m/s, two orders smaller than the speed of light. A significant plasmonic Doppler effect can emerge in electrically biased graphene when the electron drift velocity $v_d$ reaches a substantial fraction of the plasmon velocity[18,23]. Such a Doppler effect has been predicted to break the time reversal symmetry in the graphene optical response in the nonlocal limit and create non-reciprocal surface plasmon propagations[24-31].

Here we report the first experimental observation of the plasmonic Doppler effect in monolayer graphene. Taking advantage of the cryogenic scanning near-field infrared nanoscopy, we can measure the Doppler-induced wavelength shift in real space even under a large bias current in graphene. Our two-terminal graphene device is composed of ultraclean monolayer graphene fully encapsulated in hexagonal boron nitride (hBN) and a nanofabricated gold nanobar as the integrated plasmon launcher. The plasmon launched by the gold nanobar is imaged by the near-field infrared nanoscopy, and it exhibits significant modulation by the electrical bias current (and therefore the electron drift velocity). We quantify the plasmonic Doppler effect by monitoring the plasmon wavelength change for positive and negative electrical current and observe a Doppler-induced wavelength modulation as large as 3.6% for a bias current density of $\pm\,0.8$ mA/μm. Our experimental results agree well with the existing theoretical model[30]. This strong plasmonic Doppler effect opens new doors for exploring nonequilibrium plasmons[32-35] and non-reciprocal plasmonic[36-38] phenomena in graphene and other high mobility 2D materials.

**Doppler-induced wavelength shift**

Figure 1a displays the schematic of our device fabricated on a $SiO_2$/Si substrate. Ultraclean monolayer graphene was fully encapsulated by two hBN flakes using the dry-transfer technique (details of sample fabrication in Methods). Source-drain electrodes with low contact resistance are fabricated using the one-dimensional edge contact method[39]. The narrow gold nanobar (~ 500 nm in width and 9.5 μm in length) in the middle of the device is used to excite the plasmon in the graphene sheet (5 μm in width and 15 μm in length). In order to access to the graphene plasmon efficiently, we use a very thin top hBN layer with a thickness of ~ 2 nm. Figure 1b illustrates the Doppler effect of graphene plasmon launched by the gold nanobar under positive and negative carrier flow directions. The sharp edge of the gold nanobar serves as an efficient launcher of the plasmon in graphene[12,20], which propagates away from the nanobar. The presence of a carrier flow breaks time reversal symmetry ($\omega_{kx} \neq \omega_{-kx}$) and results in different plasmonic group velocities depending on the direction of the external driving current[26]. This leads to a stretched (compressed) plasmon wavelength for plasmons propagating along (against) the carrier flow direction. In Figure 1b, the wavelength change is exaggerated for better illustrating the idea.

We probe the graphene plasmon with a home-built near-field infrared nanoscopy setup at cryogenic temperature (details in the Methods). The base temperature of the sample is fixed at 25 K, which helps to dissipate the unintended Joule heating caused by the large current through the graphene channel. In addition, the phonon scattering of the graphene plasmon is strongly suppressed at low temperature, resulting in a higher plasmon quality factor[3,12] and thus a better accuracy in determining the plasmon wavelength. To probe the graphene plasmon, a 10.6 µm $CO_2$ laser was focused on the sample by an aspheric ZnSe lens with a spot size around 10 µm. This guarantees a uniform light illumination on the nanobar and atomic force microscope (AFM) tip when scanning around the bar area (within 1 µm range), with negligibly small intensity and phase inhomogeneity. Basically, the graphene plasmon excited by the gold nanobar propagates away from the bar edge. The propagating plasmon is later scattered by the tip apex, which interferes with a background scattered light and gets detected by a HgCdTe (MCT) detector in the far field. By scanning along the plasmon propagating direction and recording the near-field interference fringes, we can determine the plasmon wavelength accurately[12,19-21,40]. Figure 1c shows the gate-dependent resistance of a representative Doppler device at 25 K. This specific device has a 2 µm thick $SiO_2$ dielectrics to minimize the doping change induced by the bias voltage ($V_b$) required in the Doppler measurements. Only the hole doping data is measured because such device with thick $SiO_2$ dielectrics breaks down easily under a high positive backgate voltage $V_g$ (See Methods). The device resistance decreases quickly with increased doping away from the charge neutral point at $V_g = -380$ V. The small resistance peak at $V_g = -780$ V is presumably due the graphene/boron nitride moiré superlattice. There is little hysteresis between the forward (blue trace) and backward gate voltage scan (red trace). Figure 1d shows the channel current of the same device as a function of $V_b$ at $V_g = -1050$ V, corresponding to a carrier density of $|n| = 7.2 \times 10^{12}$ cm$^{-2}$ based on the capacitance model (see Methods). Doppler measurements are carried out at this doping, where high quality plasmons at mid-infrared wavelengths are present. The two-terminal resistance at high doping is dominated by the electrical contact. Using the one-dimensional edge contact, we achieve a low contact resistance of ~ 870 Ω (slope of the linear fit of Fig.1d), which allows us to drive large electrical current through the graphene device.

Figure 2 shows the near-field signal under different driving currents. The data is collected from scanning the same line along the plasmon propagating direction which is perpendicular to the nanobar. At different currents, the graphene plasmons behave quite differently not only because of the Doppler-induced wavelength shift, but also because of a temperature change due to the strong Joule heating. The temperature change has two effects. First, plasmon damping increases with temperature which leads to a decreased near-field signal in amplitude; second, thermal expansion and other effects can modify the background light scattering and thus modulate the near-field interference signal. In order to eliminate those complexities induced by the thermal effects, we directly compare the plasmon wavelengths for positive and negative currents of the same magnitude, which has the same thermal load in the device but generates opposite Doppler shift. Figure 2a shows the 2D near-field signal images obtained from stacking 30 consecutive scans of a single line at + 2 mA and - 2 mA. Figure 2d shows the line profile obtained by averaging 30 scans to improve the signal-to-noise ratio. The wavelength at + 2 mA is stretched by 1.2% compared with the - 2 mA case, which is determined by averaging the wavelength shifts at the interference peaks and dips in the line profile in Fig. 2d. Figures 2b and 2e show the measured near-filed signals when the current increases to about ± 3 mA. The difference between positive and negative currents induced shift becomes more prominent. We observe a total ~ 2.2% Doppler-induced wavelength shift between + 3 mA and - 3 mA. The largest wavelength shift we can observe reaches ~ 3.6% between + 4 mA and - 4 mA (Figs. 2c and 2f). At currents higher than ± 4 mA, the plasmons become very weak and are beyond the detection limit of our setup, presumably due to the significant thermal heating. A similar Doppler shift has been observed in other graphene devices (See the Methods). In principle, Doppler-induced wavelength shift can be larger at low carrier density, because a larger carrier drift velocity can be realized at low doping with the same thermal load. Experimentally, however, the graphene plasmon becomes quite weak at low carrier doping and accurate measurements of the plasmons are very challenging with 10.6 μm laser excitation.

**Non-uniform doping effects**

Next, we rule out the possibility of doping-induced wavelength shift in our experimental observation. In order to achieve large carrier drift velocity, the applied bias ($V_b$) is on the order of few Volts. The switching of the current direction by switching the sign of $V_b$ will result in some changes in the local graphene doping level, which in turn gives rise to a finite wavelength shift. Figure 3a illustrates the $V_b$-induced doping change in graphene channel with positive and negative $V_b$. In some devices, we used $SiO_2$/Si substrate with a 2 μm thick oxide layer to significantly reduce this side effect. We experimentally investigate the graphene near-field response with changing the $V_g$ (as schematically illustrated in Fig. 3b). The graphene channel is fixed at - 3 mA while the $V_g$ is varied by adding an offset of ± 2.66 V (the bias voltages to achieve about ± 3 mA current) to - 1050 V. Figure 3c and 3d show the near-field infrared nanoscopy data of the graphene plasmons with an offset backgate voltage of - 2.66 V and + 2.66 V, respectively. Figure 3e displays the line profiles of the near-field signal with - 3 mA bias current and an offset voltage of 0 V (blue trace), -2.66 V (green trace), and 2.66 V (black trace), which are compared to the near-field signal with + 3 mA and zero offset voltage (red trace). The wavelength shift between ± 2.66 V offset gate conditions at − 3 mA (green and black traces) is 0.34%, and this gives a gate induced wavelength shift of 0.06% per Volt, which is consistent with the theoretical predicted value of 0.07% per Volt for our device conditions[20]. In the middle of the graphene channel where we performed the measurements, the gate voltage change due to the reverse of the current directions between ± 3 mA is around 2.66 V, and this corresponds to a gate induced wavelength shift of 0.17%. Therefore the doping induced wavelength change is over an order of magnitude smaller than the measured Doppler induced shift (~ 2.2% in Figs. 2b and 2e). We note that the doping effect for $Si/SiO_2$ devices with 285nm $SiO_2$ will be stronger, but the Doppler wavelength shift will still be significantly larger than the doping induced wavelength change.

**Nonequilibrium plasmon response**

The nonequilibrium plasmon response of the current-carrying monolayer graphene can be calculated using the linear response theory[30]. The density-density response function of current-carrying states can be analytically approximated by employing the random phase approximation[30]. The analytical form of graphene plasmon under a biased condition can then be expressed as $\omega_{pl}(k) = \sqrt{\frac{2D_0 W(\beta)}{\varepsilon \beta}} \sqrt{k} \left[ 1 + \gamma \frac{\sqrt{2\beta W(\beta)}}{4} \sqrt{\frac{k}{k_{TF}}} + \frac{12 - 16\alpha_{ee}^2 - 3\beta W(\beta)}{16} \frac{k}{k_{TF}} \right]$, $W(\beta) = 2\frac{1-\sqrt{1-\beta^2}}{\beta}$, $\alpha_{ee} = \frac{e^2}{\varepsilon \hbar v_F}$. Here $\gamma = \pm 1$ denotes plasmon propagation along and against the carrier flow direction,

$\beta = \frac{v_d}{v_F}$ is the normalized electron drift velocity relative to Fermi velocity with $v_F = 0.85 \times \frac{10^6 \text{m}}{\text{s}}$.

$D_0 = \frac{e^2 E_F}{\hbar^2}$ is the Drude weight of noninteracting 2D massless Dirac fermions expressed in terms of the electron charge e, Fermi energy $E_F$, and reduced Planck constant $\hbar$. $k_{TF} = 4\alpha_{ee} k_F$ is the Thomas-Fermi screening wave vector at $T = 0$ K (Ref.[41]), and $\varepsilon$ and $k_F$ are the effective dielectric constant of hBN and the Fermi wave vector. The drift electrons break the time-reversal symmetry and makes the graphene plasmon propagation non-reciprocal (Fig. 4a). In the absence of the drift current (i.e. $v_d = 0$), the $\omega \sim k$ dispersion curve is formed by two symmetric branches (blue curves in Figure 4(a)) corresponding to two counter propagating waves. The two branches are linked by $\omega_{kx} = \omega_{-kx}$, in agreement with the reciprocity and parity symmetries of the system. In contrast, with the drift current flow (Fig. 4b), there is an evident symmetry breaking of the SPPs dispersion such that the positive and negative directions become nondegenerate. The forward current (upstream) lifts the dispersion and the backward current (downstream) depresses the dispersion resulting in a wavelength shift in the plasmon. The asymmetry of the dispersion curve becomes more prominent when the drift velocity increases. Figure 4c displays our experimental data of plasmon wavevector shift between positive and negative electrical current (symbols) at different ratios of drift velocity over plasmon velocity ($\frac{v_d}{v_p}$). The green line in Figure 4c shows the theoretical prediction, which is obtained by calculating the difference between the upstream and downstream dispersion curves in Fig. 4b. The experimental values of $v_d$ are obtained using the equation $v_d = \frac{J}{ne} = \frac{I}{new}$, where J is the current density, I is the driving current, $w = 5$ μm is the width of the graphene device, and the carrier density $|n| = 7.2 \times 10^{12}$ cm$^{-2}$. The plasmon velocity $v_p$ is directly obtained from the measured plasmon wavelength $\lambda_p$ with the relation $v_p = \lambda_p \times f$, where $\lambda_p$ is the plasmon wavelength and f = 28.3 THz is the probing laser frequency. At a driving current of $\pm$ 3 mA, switch of the current direction yields a total carrier drift velocity change of ~ $1.04 \times 10^5$ m/s, corresponding to ~ 2.4% of the plasmon velocity. The observed Doppler induced shift reaches ~ 2.2%. Our experimental observations are consistent with the theoretical predictions as shown in Fig. 4c.

To achieve stronger Doppler shift and non-reciprocal propagation, we need to further increase the carrier drift velocities [24,26-30] with higher bias current. This can potentially be achieved by implementing more effective heat sink and by using double-layer graphene[42]. Another approach is to use short electrical driving pulses, which has much lower duty cycle and can avoid undesirable Joule heating effects. These future improvements can significantly increase the achievable carrier drift velocity and enhance the Fizeau drag effects in graphene, which will open new opportunities to study unidirectional plasmonic phenomena and highly nonequilibrium plasmons.

# References:

1. Foucault, L. Méthode générale pour mesurer la vitesse de la lumière dans l'air et les milieux transparents. Vitesses relatives de la lumière dans l'air et dans l'eau. Projet d'expérience sur la vitesse de propagation du calorique rayonnant. CR Hebd. *Seances Acad. Sci* **30**, 551-560 (1850).
2. Christensen, T. in *From Classical to Quantum Plasmonics in Three and Two Dimensions* 13-35 (Springer, 2017).
3. Maier, S. A. *Plasmonics: fundamentals and applications*. (Springer Science & Business Media, 2007).
4. Jablan, M., Soljačić, M. & Buljan, H. Plasmons in graphene: fundamental properties and potential applications. *Proceedings of the IEEE* **101**, 1689-1704 (2013).
5. Low, T. *et al.* Polaritons in layered two-dimensional materials. *Nature materials* **16**, 182-194 (2017).
6. Novotny, L. & Hecht, B. *Principles of nano-optics*. (Cambridge university press, 2012).
7. Kauranen, M. & Zayats, A. V. Nonlinear plasmonics. *Nature photonics* **6**, 737 (2012).
8. Bozhevolnyi, S. I., Martin-Moreno, L. & Garcia-Vidal, F. *Quantum plasmonics*. (Springer, 2017).
9. Yu, N. & Capasso, F. Flat optics with designer metasurfaces. *Nature materials* **13**, 139-150 (2014).
10. Xia, F., Wang, H., Xiao, D., Dubey, M. & Ramasubramaniam, A. Two-dimensional material nanophotonics. *Nature Photonics* **8**, 899 (2014).
11. Neto, A. C., Guinea, F., Peres, N. M., Novoselov, K. S. & Geim, A. K. The electronic properties of graphene. *Reviews of modern physics* **81**, 109 (2009).
12. Ni, G. *et al.* Fundamental limits to graphene plasmonics. *Nature* **557**, 530 (2018).
13. Koppens, F. H., Chang, D. E. & Garcia de Abajo, F. J. Graphene plasmonics: a platform for strong light–matter interactions. *Nano letters* **11**, 3370-3377 (2011).
14. Dean, C. R. *et al.* Boron nitride substrates for high-quality graphene electronics. *Nature nanotechnology* **5**, 722-726 (2010).
15. Mak, K. F. *et al.* Measurement of the optical conductivity of graphene. *Physical review letters* **101**, 196405 (2008).
16. Zhang, Y., Small, J. P., Pontius, W. V. & Kim, P. Fabrication and electric-field-dependent transport measurements of mesoscopic graphite devices. *Applied Physics Letters* **86**, 073104 (2005).
17. Yu, Y.-J. *et al.* Tuning the graphene work function by electric field effect. *Nano letters* **9**, 3430-3434 (2009).
18. Dorgan, V. E., Bae, M.-H. & Pop, E. Mobility and saturation velocity in graphene on $SiO_2$. *Applied Physics Letters* **97**, 082112 (2010).
19. Woessner, A. *et al.* Highly confined low-loss plasmons in graphene–boron nitride heterostructures. *Nature materials* **14**, 421 (2015).
20. Fei, Z. *et al.* Gate-tuning of graphene plasmons revealed by infrared nano-imaging. *Nature* **487**, 82 (2012).
21. Chen, J. *et al.* Optical nano-imaging of gate-tunable graphene plasmons. *Nature* **487**, 77-81 (2012).
22. Jablan, M., Buljan, H. & Soljačić, M. Plasmonics in graphene at infrared frequencies. *Physical review B* **80**, 245435 (2009).
23. Ramamoorthy, H. *et al.* "Freeing" Graphene from Its Substrate: Observing Intrinsic Velocity Saturation with Rapid Electrical Pulsing. *Nano letters* **16**, 399-403 (2015).
24. Borgnia, D. S., Phan, T. V. & Levitov, L. S. Quasi-Relativistic Doppler Effect and Non-Reciprocal Plasmons in Graphene. *arXiv preprint arXiv:1512.09044* (2015).
25. Correas-Serrano, D. & Gomez-Diaz, J. Non-Reciprocal and Collimated Surface Plasmons in Drift-biased Graphene Metasurfaces. *arXiv preprint arXiv:1905.10699* (2019).
26. Morgado, T. A. & Silveirinha, M. G. Drift-induced unidirectional graphene plasmons. *ACS Photonics* **5**, 4253-4258 (2018).
27. Sabbaghi, M., Lee, H.-W. & Stauber, T. Electro-optics of current-carrying graphene. *Physical Review B* **98**, 075424 (2018).
28. Sabbaghi, M., Lee, H.-W., Stauber, T. & Kim, K. S. Drift-induced modifications to the dynamical polarization of graphene. *Physical Review B* **92**, 195429 (2015).
29. Wenger, T., Viola, G., Kinaret, J., Fogelström, M. & Tassin, P. Current-controlled light scattering and asymmetric plasmon propagation in graphene. *Physical Review B* **97**, 085419 (2018).

30  Van Duppen, B., Tomadin, A., Grigorenko, A. N. & Polini, M. Current-induced birefringent absorption and non-reciprocal plasmons in graphene. *2D Materials* **3**, 015011 (2016).
31  Song, J. C. & Rudner, M. S. Chiral plasmons without magnetic field. *Proceedings of the National Academy of Sciences* **113**, 4658-4663 (2016).
32  Hamm, J. M., Page, A. F., Bravo-Abad, J., Garcia-Vidal, F. J. & Hess, O. Nonequilibrium plasmon emission drives ultrafast carrier relaxation dynamics in photoexcited graphene. *Physical Review B* **93**, 041408 (2016).
33  Morgado, T. A. & Silveirinha, M. G. Negative Landau damping in bilayer graphene. *Physical review letters* **119**, 133901 (2017).
34  Ni, G. *et al.* Ultrafast optical switching of infrared plasmon polaritons in high-mobility graphene. *Nature Photonics* **10**, 244 (2016).
35  Page, A. F., Ballout, F., Hess, O. & Hamm, J. M. Nonequilibrium plasmons with gain in graphene. *Physical Review B* **91**, 075404 (2015).
36  Lin, X. *et al.* Unidirectional surface plasmons in nonreciprocal graphene. *New Journal of Physics* **15**, 113003 (2013).
37  Wang, Z., Chong, Y., Joannopoulos, J. D. & Soljačić, M. Reflection-free one-way edge modes in a gyromagnetic photonic crystal. *Physical review letters* **100**, 013905 (2008).
38  Yu, Z., Veronis, G., Wang, Z. & Fan, S. One-way electromagnetic waveguide formed at the interface between a plasmonic metal under a static magnetic field and a photonic crystal. *Physical review letters* **100**, 023902 (2008).
39  Wang, L. *et al.* One-dimensional electrical contact to a two-dimensional material. *Science* **342**, 614-617 (2013).
40  Ocelic, N., Huber, A. & Hillenbrand, R. Pseudoheterodyne detection for background-free near-field spectroscopy. *Applied Physics Letters* **89**, 101124 (2006).
41  Kotov, V. N., Uchoa, B., Pereira, V. M., Guinea, F. & Neto, A. C. Electron-electron interactions in graphene: Current status and perspectives. *Reviews of Modern Physics* **84**, 1067 (2012).
42  Morgado, T. A. & Silveirinha, M. G. Nonlocal effects and enhanced nonreciprocity in current-driven graphene systems. *Physical Review B* **102**, 075102 (2020).
43  Geick, R., Perry, C. & Rupprecht, G. Normal modes in hexagonal boron nitride. *Physical Review* **146**, 543 (1966).
44  Hwang, C. *et al.* Fermi velocity engineering in graphene by substrate modification. *Scientific reports* **2**, 590 (2012).

**Acknowledgements**: The device fabrication and characterization and theoretical analysis of the work is supported by the Director, Office of Science, Office of Basic Energy Sciences, Materials Sciences and Engineering Division of the US Department of Energy under contract number DE-AC02-05CH11231 (sp2-Bonded Materials Program KC2207). The cryogenic near-field nanoscopy measurement was supported by the NSF award 1808635. K.W. and T.T. acknowledge support from the Elemental Strategy Initiative conducted by the MEXT, Japan and the CREST (JPMJCR15F3), JST.

**Author contributions:** F.W. conceived the research. W.Z. and S.Z. carried out near-field optical measurements. W.Z., S.Z., S.W., S. Y., F.W. performed data analysis. W.Z., S.Z., H.L., S.W., M.I.B.U, S.K., Y.J., X.X. fabricated the graphene devices. K.W. and T.T. grew hexagonal boron nitride crystals. All authors discussed the results and wrote the manuscript.

**Author information:** The authors declare no competing interests. Correspondence and requests for materials should be addressed to fengwang76@berkeley.edu.

**Data availability:** The data that support the findings of this study are available from the corresponding author upon reasonable request.

**Main figure legends:**

**Fig. 1 | Schematic view of Doppler effect in a graphene device.** (**a**) Schematics of the plasmonic Doppler device, which includes hBN encapsulated graphene, a top gold nanobar as plasmon launcher, and source-drain electrodes with one-dimensional edge contacts for electrically driving the current. (**b**) Illustration of the plasmonic Doppler effect in graphene. The sharp edge of the gold nanobar efficiently excites plasmons propagating away from the gold nanobar. The wavelength of propagating plasmons will be shifted due to the drifting electron medium. Plasmons propagating along with the drifting electrons will have an enhanced speed and longer wavelength, while plasmons propagating against the drifting electrons will have a shorter wavelength. (**c**) Gate-dependent two-terminal resistance of the graphene device. The resistance decreases quickly with increased carrier doping in the device. The second resistance peak corresponds to the second Dirac point due to unintentional alignment of the graphene and hBN lattice. (**d**) I-V curve of the Doppler device at $V_g$ = - 1050 V. The two-terminal resistance is about 870 Ω, indicating the high quality of the electrical contact. The inset of (**d**) shows the optical image of the graphene device. Due to the thick $SiO_2$ layer, the contrasts of the very thin top hBN and graphene are very weak. The graphene channel is indicated by the red shaded region. The scale bar in the inset of (**d**) is 10 μm.

**Fig. 2 | Near-field signal of the propagating plasmons under different driving currents.** Near-field data at (**a**) + 2 mA and - 2 mA, (**b**) + 3 mA and - 3 mA, and (**c**) + 4 mA and - 4 mA. Data of 30 consecutive scans along the same line on the sample is shown in **a-c**. The gold nanobar locates at the left in **a-c** and the graphene plasmons are launched and propagate to the right. The red and blue arrows in **a-c** indicate the carrier flow directions. (**d**)-(**f**) are the corresponding line profiles averaged over the 30 scans shown in **a-c**. During the measurement, the AFM is controlled to always scan the same line on the sample while recording the third order harmonics of the near-field signal. The amplitude of the near-field signal degrades at large current, presumably due to Joule heating in the device. The Doppler effect induces a wavelength increase of 1.2%, 2.2%, and 3.6% for 2 mA, 3 mA, and 4 mA relative to their negative current counterpart, respectively. The wavelength shift is estimated by averaging the wavelength shifts for consecutive interference extrema positions in the near-field line profiles shown in d-f.

**Fig. 3 | Gating dependence of graphene plasmon wavelength.** (**a**) Illustration of the non-uniform doping of the Doppler device at ± 3mA bias current under a fixed $V_g$ = - 1050 V. The corresponding bias voltage is 2.66V. The middle area of the graphene experience different effective gating voltages for positive bias and negative bias. As shown in a, a negative $V_b$ results in a decreased hole doping while a positive $V_b$ leads to an increased hole doping. (**b**) Illustration of the control measurements to determine the gating induced plasmon wavelength change under a driving current. The electrical current in graphene is kept at - 3 mA, while the backgate voltage is varied by ± 2.66 V. (**c**) The near-field infrared nanoscopy data of the graphene plasmons with an offset backgate voltage of - 2.66 V. (**d**) The near-field data at an offset backgate voltage of + 2.66 V. The blue arrows in c and d indicate the carrier flow directions. (**e**) Line profiles of the near-field signal with - 3 mA bias current and an offset voltage of 0 V (blue trace), -2.66 V (green trace), and 2.66 V (black trace), which are compared to the near-field signal with +3 mA bias current and zero offset voltage (red trace). It shows that the doping-induced wavelength change is negligible than the experimentally observed Doppler-induced shift in the device.

**Fig. 4 | Graphene plasmon dispersion and Doppler-induced wavelength shift.** (**a**) Plasmon dispersion calculated from the linear response function theory under different carrier drift velocities[30]. The drift carrier velocity dramatically modifies the dispersion of the plasmon and results in non-reciprocal plasmon propagation. For $v_d$ > 0 ($v_d$ < 0), the carriers propagate parallel (antiparallel) to the plasmons. Plasmons propagating with the drift are blue-shifted for a given wavevector and have a higher group velocity. Conversely, the plasmons propagating antiparallel to the drift velocity are red-shifted and have a lower group velocity. These effects increase with higher electron drift velocity. (**b**) Wavelength shift of the graphene plasmon under different carrier drift velocity for the downstream and upstream branches ($v_F$ = 0.85 × 10$^6$ m/s, $\varepsilon$ = 5.45)[43,44]. The plasmon energy is set the same as the laser probing energy (~ 117 meV). Two branches show very different behavior at high drift velocity side. (**c**) Experimental data of the plasmon wavevector shift between the positive and negative bias current (symbols) and the corresponding theoretical prediction (green line) at different electron drift velocities. The experimental error bar is determined from the statistics of the wavelength shift calculated at different extrema of the near-field signal in Fig. 2.

**Methods:**

Encapsulated graphene device for near-field infrared nanoscopy measurements

We used a dry transfer method with a propylene carbonate (PPC) stamp to fabricate the hBN encapsulated graphene sample. Thin hBN (2 nm), monolayer graphene, and thick layer of hBN were first exfoliated onto Si substrates with a 285 nm $SiO_2$ layer. We then used a PPC stamp to pick up the thick layer hBN, graphene, and the thin hBN in sequence to fully encapsulate the graphene channel. In order to get an ultra-clean surface, the PPC stamp with the above heterostructure was then flipped over and stamped onto a clean $SiO_2$/Si substrate with either 2 μm or 285 nm $SiO_2$ dielectrics thickness. Encapsulated graphene samples were made into devices with one-dimensional edge contacts. In brief, standard e-beam lithography (EBL) was used to open graphene contact windows on the PMMA-coated (PMMA 950 A4) sample substrates. Reactive ion etching (RIE) with $CHF_3$ and $O_2$ etching gases (40 sccm and 6 sccm) was used to etch hBN to expose graphene edges with saw-tooth edge shapes. A separate EBL process was then carried out to design the electrode patterns of a two-terminal device and a narrow nanobar structure in the middle of graphene. Immediately before the metal deposition, the samples were treated with mild oxygen plasma to expose a clean graphene edges. Cr/Au (typically 5 nm/75 nm) electrode was made using an e-beam evaporator equipped with a water cooling system at high vacuum ($< 1 \times 10^{-6}$ Torr). The surfaces of the devices were cleaned by a mild hydrogen plasma treatment (250 °C, 10 sccm $H_2$) for about 30 min. The treated devices were kept in a conductive case for 1 or 2 days to avoid potential charging issue before any near-field infrared nanoscopy measurements.

Cryogenic near-field infrared nanoscopy measurements

Our cryogenic near-field infrared nanoscope was based on a home-made AFM which has the capability to work at high vacuum and low temperature. The whole AFM setup was built inside a closed cycle cryostat, and the AFM head was connected to the cold plate by a soft copper braid in order to damp the vibration from the pulse tube. The lowest sample temperature achieved in our AFM system was 25 K. A $CO_2$ laser was coupled into the vacuum chamber through an aspheric ZnSe lens with 0.45 NA. The position of the lens was controlled by a vacuum compatible stage. The back scattered light from the tip was collected by MCT in a self-homodyne configuration, the near-field signal was demodulated at the third harmonic of the tapping frequency to suppress the background. In order to increase the signal to noise ratio, the time constant was set to be 10 mS. During the scanning, the turbo pump was turned off in order to minimize the mechanical vibration. The background vacuum level remains below $1 \times 10^{-6}$ mbar through all the measurements.

Estimation of carrier density in graphene

The dielectrics in the plate capacitor formed between graphene and Si backgate were composed of the $SiO_2$ dielectrics (medium 1) and the bottom hBN layer (medium 2). The total geometric capacitance per unit area was calculated from the expression $\frac{1}{C_{total}} = \frac{1}{C_1} + \frac{1}{C_2}$, where $C_1 = \epsilon_0 \epsilon_{r1}/d_1$ and $C_2 = \epsilon_0 \epsilon_{r2}/d_2$ are the capacitance per unit area of $SiO_2$ and hBN, $\epsilon_0$ is the vacuum permittivity, $\epsilon_{r1} = \epsilon_{r2} = 3.9$. For the device measured in Fig. 2 and Fig. 3, $d_1 = 2$ μm and $d_2 = 40$ nm. The carrier density was obtained by $n = C_{total} \times V_g$, where $V_g$ was measured relative to the charge neutrality point.

In situ electrical gating and transport measurement of the device

The very high backgate voltage used for device on 2 μm thick $SiO_2$ dielectrics could trigger gas ionization in the vacuum. In order to protect the sample from sudden discharging breakdown, we scanned the backgate voltage slowly (0.5 V/s) at hole doping side for several rounds to make sure the response of the device becomes stable (Keithley 2410). During the scanning of backgate voltage, the conductance of the device was monitored with a DC bias of 1 mV (Keithley 2614B). Electron drift velocity at a fixed backgate voltage was controlled by varying the amplitude and polarity of the DC bias (Keithley 2614B).

Doppler-induced wavelength shift in a second device

Extended Data Figs. 1-4 show the plasmonic Doppler effect in another device with 285 nm thick $SiO_2$ dielectrics. The width of the graphene channel is $w = 2.5$ μm, and the near-field measurements are performed at a carrier density of $7.0 \times 10^{12}$ $cm^{-2}$. In this device, we are able to measure high quality plasmons propagating on both sides of the gold nanobar. The graphene channel current at discreet bias voltages measured at the same doping condition and temperature is shown in Extended Data Fig. 1. At high bias condition, the resistance of two-terminal device is about 1.5 kΩ, indicating a low contact resistance for this narrow channel device.

On the left side of the gold nanobar, the negative current flow enhances the plasmon speed and results in a stretched plasmon wavelength, while the positive current flow reduces the plasmon speed with a compressed plasmon wavelength. Extended Data Fig. 2c shows the 2D images of near-field signals at ±0.4 mA where 30 consecutive scans along the same line are recorded. The gold nanobar locates at the right. The graphene plasmons are launched and propagate to the left, which is consistent with the schematics in Extended Data Figs. 2a and 2b. The positive and negative current of 0.4 mA generates a minor shift of plasmon wavelength (~ 1.1%) as shown in the line profile in Extended Data Fig. 2d (average from 30 line scans in Extended Data Fig. 2c). As we increase the current to be ± 1.2 mA, the plasmon wavelength difference becomes more prominent (~ 2.3% in Extended Data Fig. 2f). When the current reaches + 1.7 mA and - 1.9 mA, the wavelength shift becomes ~ 3.1%. The wavelength shift is extracted by averaging the shift at the interference extrema.

On the right side of the gold nanobar (Extended Data Fig. 3), the plasmon wavelength shift due to the Doppler effect is reversed. The negative current flow will generate a compressed wavelength and positive current flow will give rise to a stretched wavelength. This is clearly observed in the line profiles in Extended Data Figs. 3d and 3f which are obtained from averaging 30 consecutive line scans shown in Extended Data Figs. 3c and 3e. For the plasmons on the right side of the gold nanobar, the wavelength difference is determined to be ~ 1.5% for ± 0.4 mA current and ~ 3.1% for ± 1.2 mA current. The plasmon quality for 1.9 mA driving current (not shown) is too low for the right side to reliably determine the plasmon wavelength shift. Our experimental observations can be captured well by the theory as shown in Extended Data Fig. 4.

Breakdown of device at very high positive backgate voltage in thick oxide devices

For the 2 μm $SiO_2$ dielectric devices we found the ultra-high backgate voltage at the positive side (few hundreds V ~ 1 kV) can trigger series of gas ionization in high vacuum where the mean free paths of the residue gas molecules become several meters. This process breaks the graphene device as shown in Extended Data Fig. 5 and therefore we limit our measurements of graphene plasmon at the hole doping side by applying negative backgate voltages. Since the graphene holds particle-hole symmetry, the Doppler effect should be the same for the electron side[24,30].

Formation of graphene/hBN moiré superlattice in the device

The small resistance peak in our device as shown in Fig. 1c in the manuscript is most likely due to the formation of graphene/hBN moiré superlattice. It has been well established in transport studies of hBN-encapsulated graphene that a second small resistance peak can be observed when the graphene and hBN happen to align within a small angle ($< 2°$)[45]. This alignment can often be inferred from the alignment of straight edges between the graphene and hBN layers. To enhance the graphene and hBN contrast on thick $SiO_2$ dielectric layer, Extended Data Fig. 6 shows the optical microscope picture of our device, in which we have adjusted the contrast to observe the graphene and hBN layers. As shown in the photo, the graphene straight edge (white dashed line) nearly aligns with an edge of top hBN (yellow dashed line), which can potentially form graphene/hBN moiré superlattice with long period. The carrier density at the small resistance peak in our device is $n_s \sim 3.98 \times 10^{12}$ cm$^{-2}$, which corresponds to a moiré period of ~10.3 nm and alignment angle of ~ 0.93°[45,46].

Nonequilibrium response of the plasmons in a current-carrying graphene

The optical properties of a current-carrying graphene sheet can be calculated by using linear response theory. We follow the model in Ref. 30 to calculate the nonequilibrium plasmon response. The plasmon dispersion of monolayer graphene at various carrier drift velocities can be obtained as the roots of the real part of $\varepsilon(\mathbf{q}, \omega; 0) = 1 - \frac{2\pi e^2}{q\epsilon} \chi_{nn}^{(0)}(\mathbf{q}, \omega; 0)$, where $\chi_{nn}^{(0)}(\mathbf{q}, \omega; 0)$ is the response function at zero temperature. $\epsilon$ and $e$ are the effective dielectric constant of hBN and electron charge. In the limit $\bar{k} < \bar{\omega} \ll \mu$, the real part of the response function in equation can be expanded as

$$\frac{\text{Re}\chi_{nn}^{(0)}(\mathbf{q},\omega;0)}{D(\mu)} \approx \frac{W(\beta)}{2\beta} \frac{\bar{k}^2}{\bar{\omega}^2} - \frac{\bar{k}^2}{8} + \gamma \frac{W(\beta)^2}{4\beta} \frac{\bar{k}^3}{\bar{\omega}^3} + \gamma \frac{\beta}{32} \frac{\bar{k}^3}{\bar{\omega}} + \frac{\bar{k}^4}{8\bar{\omega}} + \frac{3W(\beta)^2}{8\beta^2} \frac{\bar{k}^4}{\bar{\omega}^4} + \cdots,$$

$$\bar{k} = \frac{\hbar v_F k}{\mu}, \quad \bar{\omega} = \frac{\hbar \omega}{\mu}, \quad W(\beta) = 2\frac{1-\sqrt{1-\beta^2}}{\beta}, \tag{1}$$

$\beta = \frac{v_d}{v_F}$ is the normalized electron drift velocity $v_d$ relative to Fermi velocity $v_F$. $\mu$ is the chemical potential of the graphene. $\gamma = +1(-1)$ denotes 'upstream' ('downstream') plasmon propagation. $D(\mu)$ is the two dimensional (2D) massless Dirac fermion (MDF) density of states. In the long wavelength limit $\bar{k} \ll 1$, the plasmon dispersion can be further reduced to

$$\omega_{pl}(k) = \sqrt{\frac{2D_0 W(\beta)}{\varepsilon \beta}} \sqrt{k} \left[ 1 + \gamma \frac{\sqrt{2\beta W(\beta)}}{4} \sqrt{\frac{k}{k_{TF}}} + \frac{12 - 16\alpha_{ee}^2 - 3\beta W(\beta)}{16} \frac{k}{k_{TF}} \right], \quad \alpha_{ee} = \frac{e^2}{\varepsilon \hbar v_F}. \tag{2}$$

$D_0 = \frac{e^2 E_F}{\hbar^2}$ is the Drude weight of noninteracting 2D MDF expressed in terms of Fermi energy

$E_F$, and reduced Planck constant $\hbar$. $k_{TF} = 4\alpha_{ee}k_F$ is the Thomas-Fermi screening wave vector at $T = 0$ K. $k_F$ is the Fermi wave vector.

**Method references:**

45    Ponomarenko, L. *et al.* Cloning of Dirac fermions in graphene superlattices. *Nature* **497**, 594-597 (2013).

46    Yankowitz, M. *et al.* Emergence of superlattice Dirac points in graphene on hexagonal boron nitride. *Nature physics* **8**, 382-386 (2012).

**Extended data figure legends:**

**Extended Data Fig. 1 | Graphene channel current at discreet bias voltages in the two-terminal device measured at 25 K at a carrier density of $7.0 \times 10^{12}$ cm$^{-2}$.**

**Extended Data Fig. 2 | Near-field signal of the propagating plasmon on the left side of the gold nanobar.** Illustration of plasmon propagation under (**a**) negative and (**b**) positive current flows. Near field data at (**c**) - 0.4 mA and + 0.4 mA, (**e**) - 1.2 mA and + 1.2 mA, and (**g**) - 1.9 mA and + 1.7 mA. (**d**), (**f**), and (**h**) are the corresponding line profiles averaged over the 30 scans. The gold nanobar locates at the right, and the graphene plasmons propagate from the right to the left.

**Extended Data Fig. 3 | Near-field signal of the propagating plasmon on the right side of the gold nanobar.** Illustration of plasmon propagation under (**a**) negative and (**b**) positive current flows. Near field data at (**c**) - 0.4 mA and + 0.4 mA, and (**e**) - 1.2 mA and + 1.2 mA. (**d**) and (**f**) are the corresponding line profiles averaged over the 30 scans. Here, the gold nanobar locates at the left, and the graphene plasmons propagate from the left to the right.

**Extended Data Fig. 4 | Comparison between the theory and experiment on the Doppler effect at different carrier drift velocities in the second device.** The width of the graphene channel is w = 2.5 µm, and the carrier density is estimated to be $|n| = 7.0 \times 10^{12}$ cm$^{-2}$.

**Extended Data Fig. 5 | Breakdown of device under high positive backgate voltages.**

**Extended Data Fig. 6 | Filtered optical image to enhance the contrast between hBN and graphene.**

Fig.1

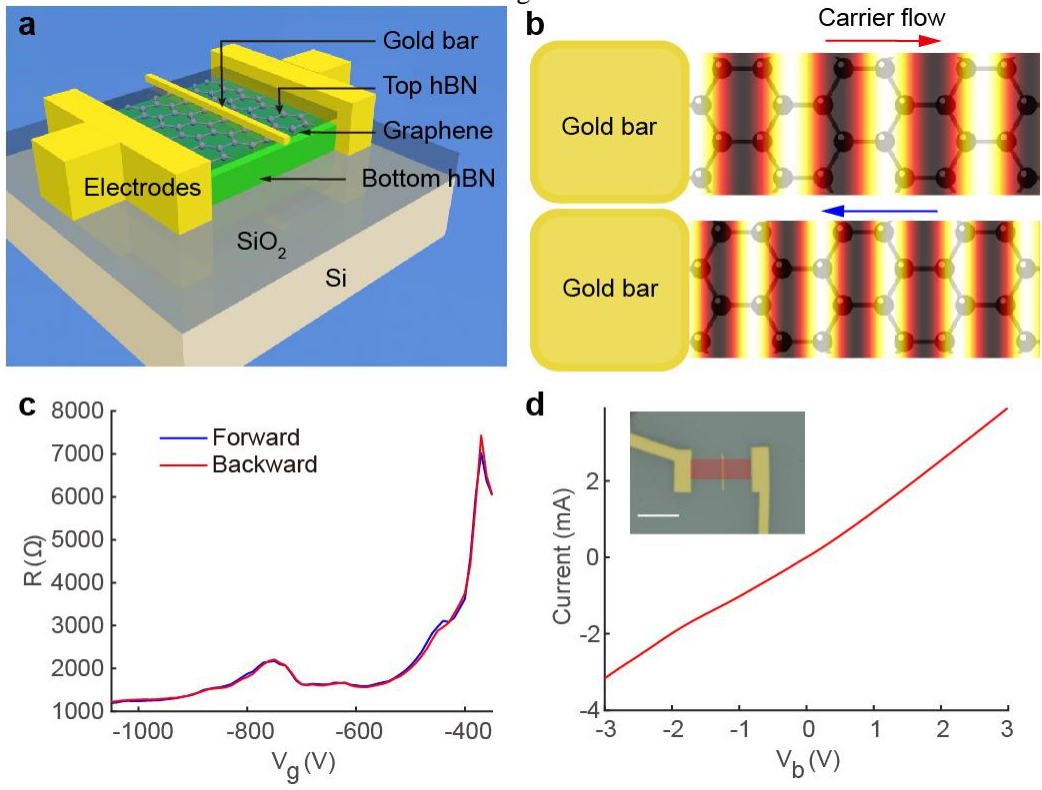

Fig.2

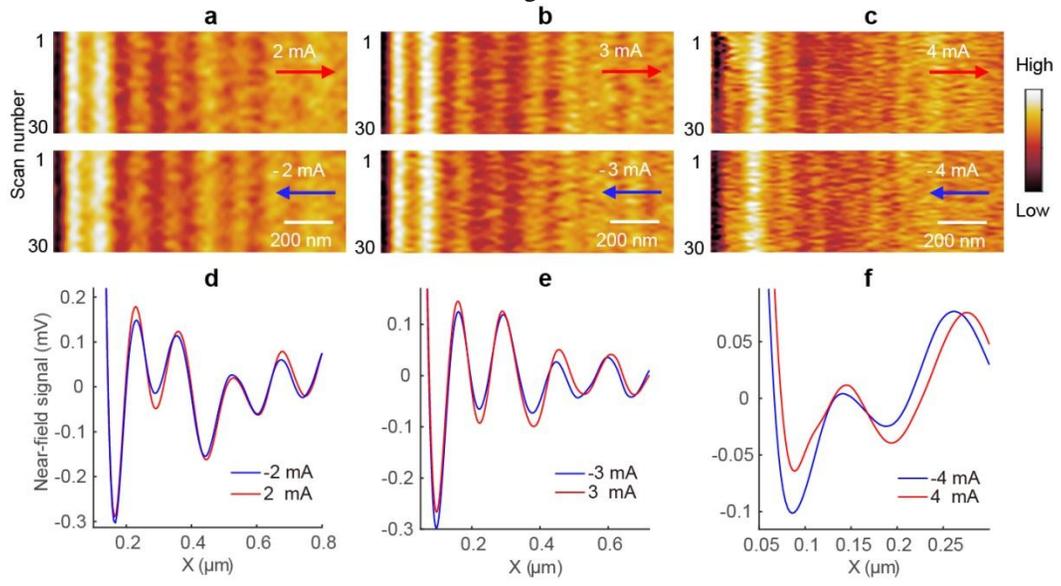

Fig. 3

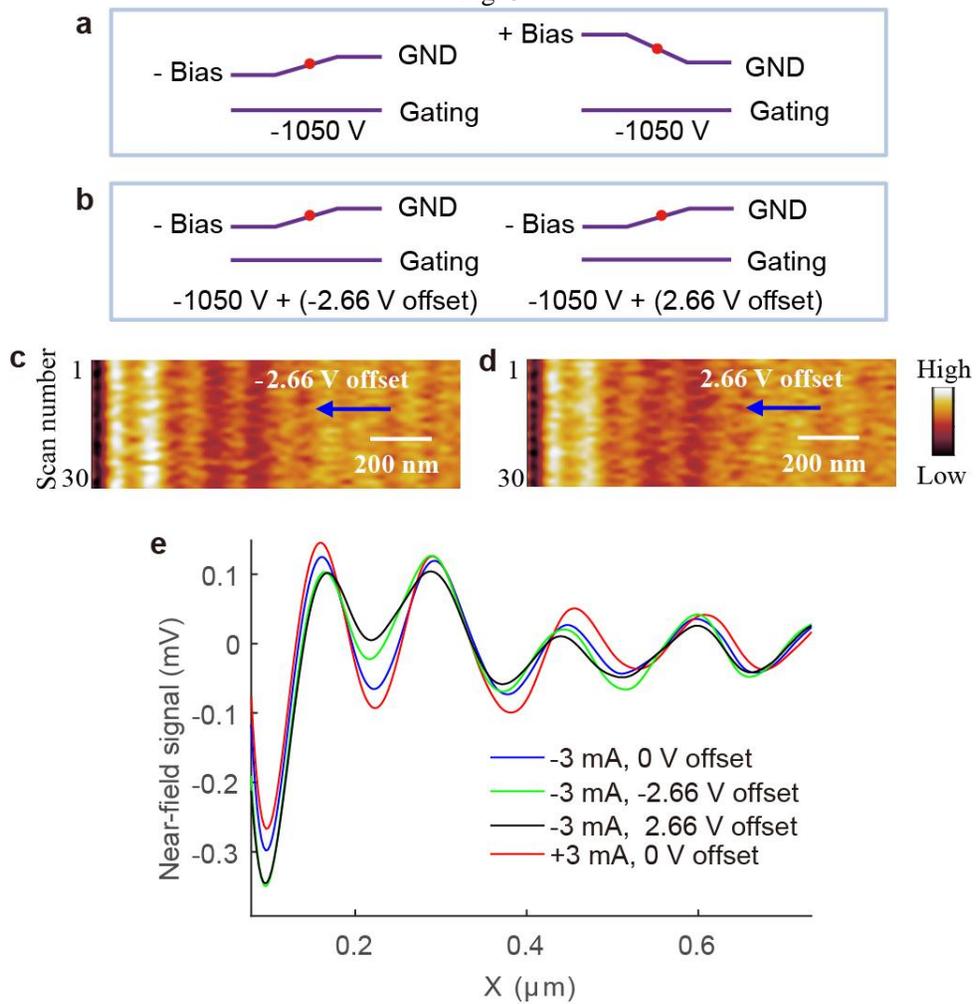

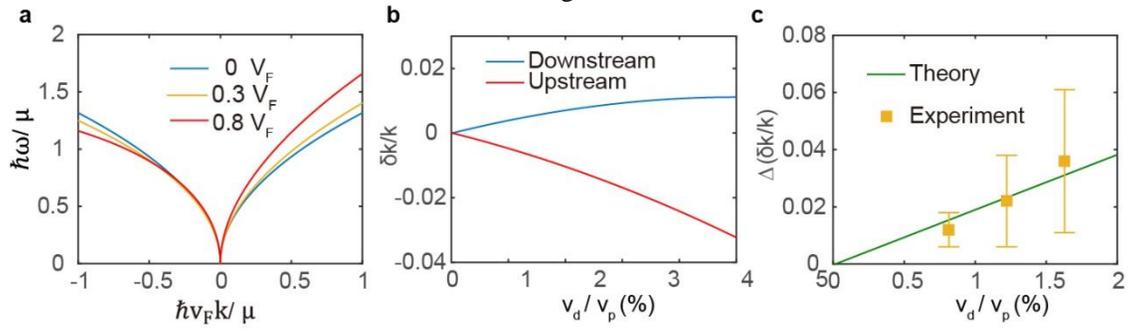

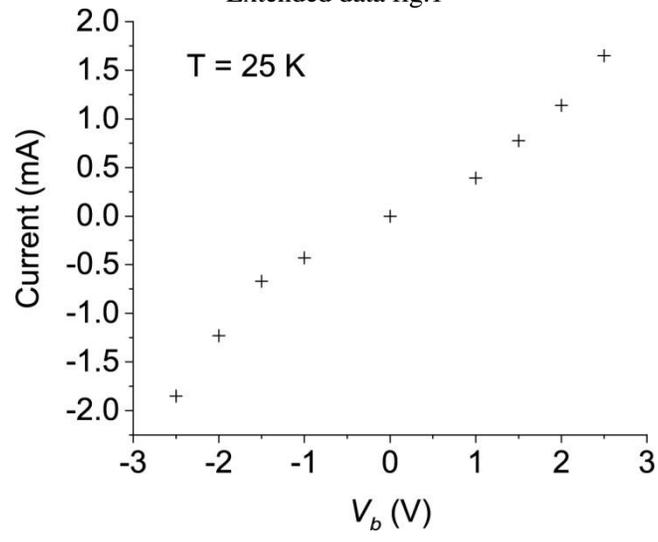

Extended data fig.1

Extended data fig.2

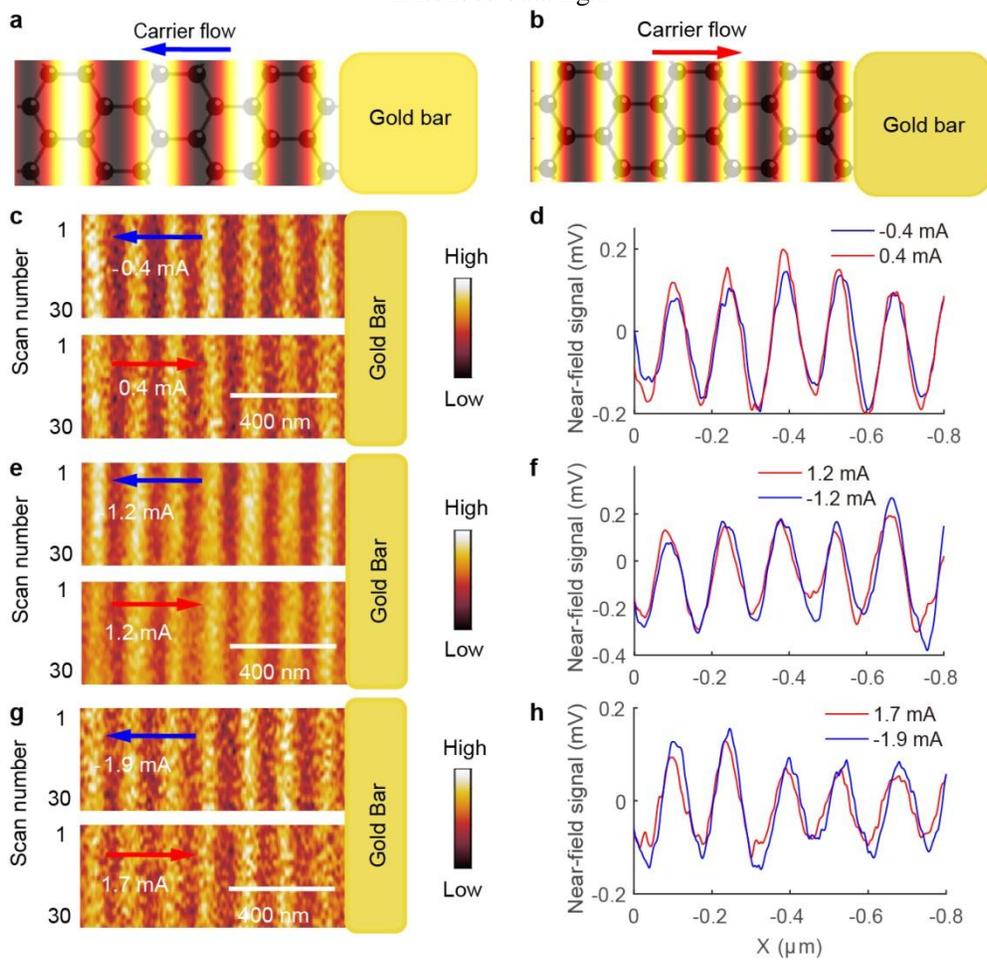

Extended data fig.3

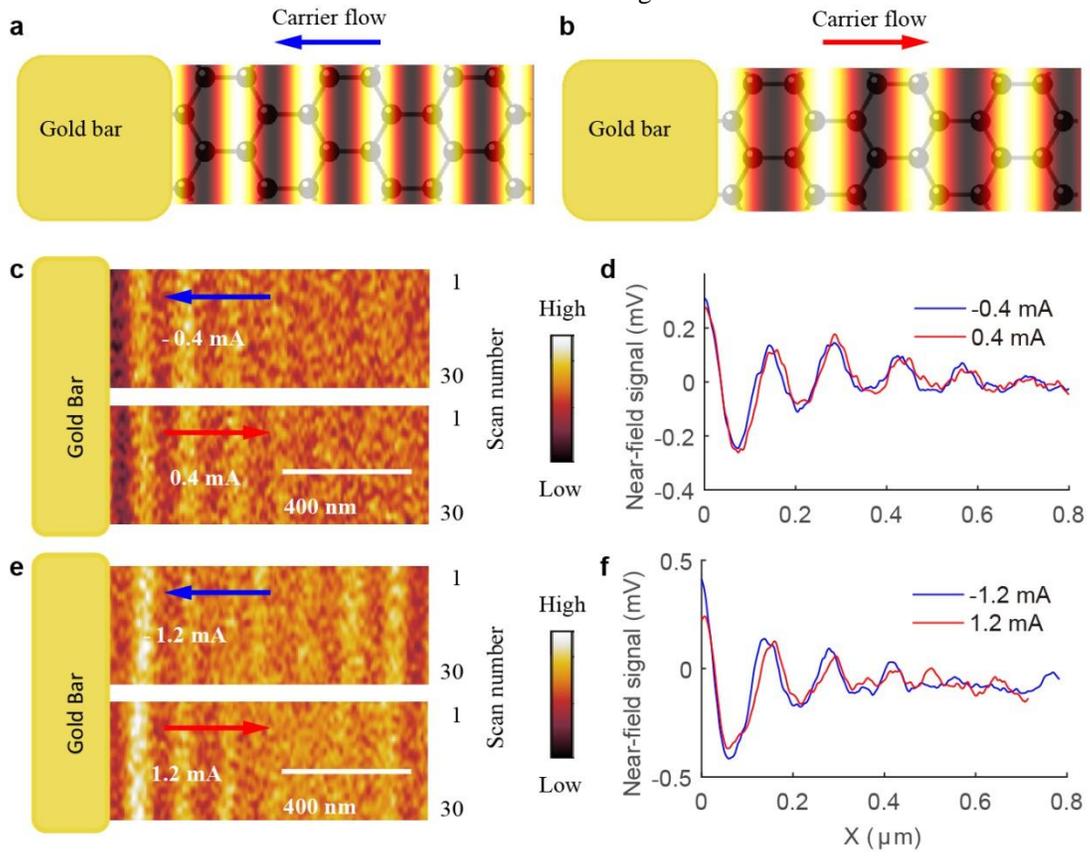

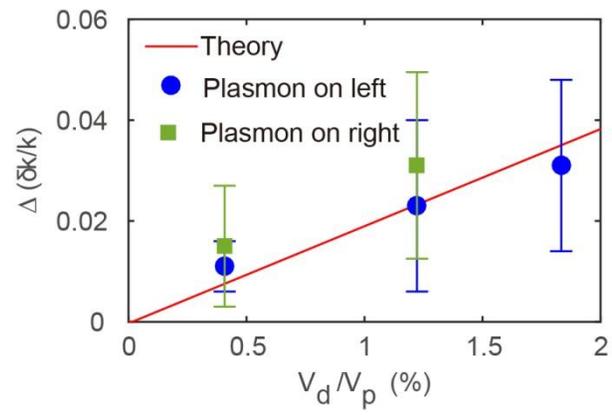

Extended data fig.4

Extended data fig.5

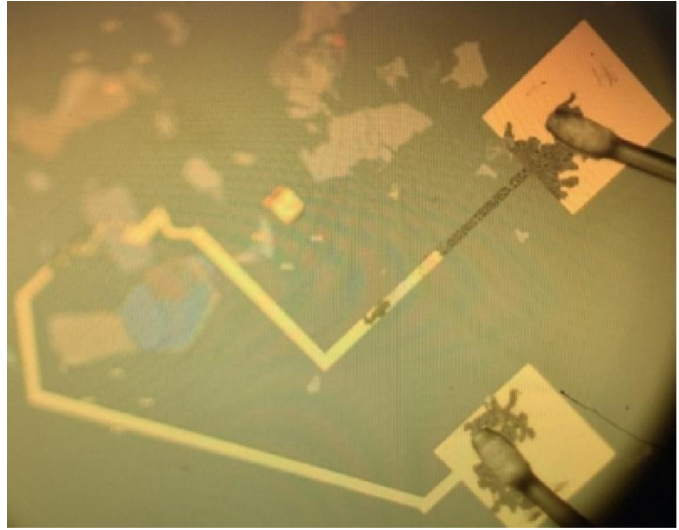

Extended data fig.6

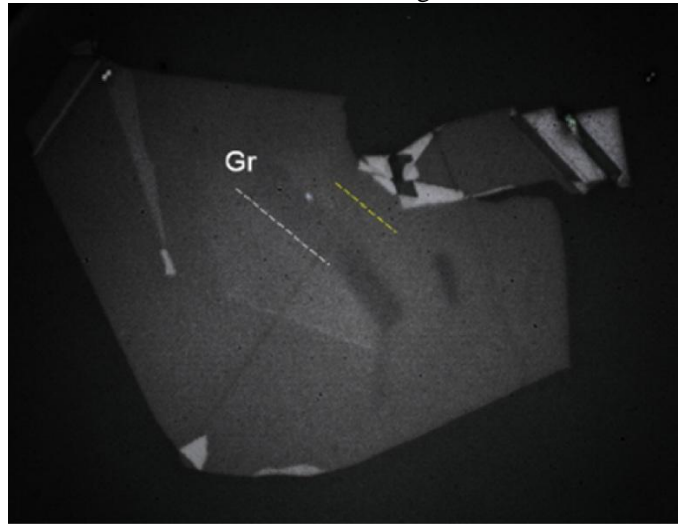